\documentclass[aps, prd, twocolumn, letterpaper, nofootinbib, superscriptaddress, balancelastpage, longbibliography,preprintnumbers]{revtex4-1}
\usepackage[dvips]{graphicx}
\usepackage{url}
\usepackage{xcolor}
\usepackage[colorlinks,citecolor=blue,urlcolor=blue,linkcolor=blue]{hyperref}
\usepackage{amsmath}
\usepackage{amssymb}

\def\openloops{\textsc{OpenLoops2}}
\def\powhegbox{\textsc{Powheg Box}}
\def\sherpa{\textsc{Sherpa}}
\def\pythia{\textsc{Pythia8}}
\def\mg5{\textsc{MG5\_aMC@NLO}}
\def\madspin{\textsc{MadSpin}}
\def\tb{\bar{t}}
\def\qb{\bar{q}}
\def\tttt{$t\bar{t}t\bar{t}$}
\def\GeV{\textrm{GeV}}
\def\lo{\textrm{LO}}
\def\nlo{\textrm{NLO}}
\def\phegas{\textsc{Helac-Phegas}}
\def\lhapdf{\textsc{LHAPDF}}

\begin{document}
\title{Hadroproduction of four top quarks in the \powhegbox{}}
\preprint{KA-TP-24-2021, MS-TP-21-30, P3H-21-085}

\author{Tom\'{a}\v{s} Je\v{z}o} 
\email{tomas.jezo@uni-muenster.de}
\affiliation{Institut f\"ur Theoretische Physik, Westf\"alische
Wilhelms-Universit\"at M\"unster, Wilhelm-Klemm-Stra\ss{}e 9, 48149 M\"unster,
Germany} 

\author{Manfred Kraus}
\email{mkraus@hep.fsu.edu}
\affiliation{Physics Department, Florida State University, Tallahassee, FL
32306-4350, U.S.A.} 

\date{\today}

\begin{abstract}
We present a new Monte Carlo event generator for the hadronic production of four
top quarks in the \powhegbox{} framework. Besides the dominant next-to-leading
order QCD corrections at $\mathcal{O}(\alpha_s^5)$ we also include all subleading
electroweak productions channels at leading-order accuracy. We validate our
theoretical predictions by comparing to parton-shower matched predictions
obtained within the MC@NLO framework for stable top quarks. Furthermore, we
investigate in detail the various sources of theoretical uncertainties. Finally,
we investigate a single lepton plus jets signature to study for the first time
the impact of the electroweak production modes as well as spin-correlation
effects at the fiducial level.
\end{abstract}

\maketitle
\section{Introduction}
Since the discovery of the top quark in $1995$ by the
CDF~\cite{Abe:1994st,Abe:1995hr} and D$\O$~\cite{D0:1995jca} experiments at the
Tevatron the landscape of top-quark phenomenology has changed dramatically. With
the advent of the Large Hadron Collider (LHC) top-quark pairs are so abundantly
produced that the top-quark physics program has entered the precision era. Due to
its high center-of-mass energy, the LHC offers the unique possibility to produce
directly final states involving multiple heavy particles. Thus, in recent years
the LHC experiments were able to measure the associated production of top-quark
pairs with gauge bosons $(W,Z,\gamma)$
\cite{Aad:2015uwa,Sirunyan:2017iyh,Sirunyan:2017uzs,Aaboud:2019njj} culminating
in the discovery of the $t\bar{t}H$ production process~\cite{Sirunyan:2018hoz,
Aaboud:2018urx}.

However, there is one rare production process that is of particular interest but
for which a discovery has not been claimed so far: the production of four top
quarks, $pp \to t\bar{t}t\bar{t}$, which can only be produced in very high
energetic collisions, due to the large threshold of $\sqrt{s}\sim 700$ GeV.
Searches for this process have been conducted by both the
ATLAS~\cite{ATLAS:2018kxv,ATLAS:2020hpj} and the CMS~\cite{CMS:2014ylv,
CMS:2017nnq,CMS:2017ocm,CMS:2019jsc,CMS:2019rvj} collaborations already for quite
some time. In the recent ATLAS measurement~\cite{ATLAS:2021kqb} of the cross
section an observed (expected) significance with respect to the background-only
hypothesis of 4.7 (2.6) standard deviations has been reported.

In recent years, a lot of attention has been devoted to the four top-quark final
state as its cross section can be significantly modified by possible physics
beyond the Standard Model (BSM)~\cite{Lillie:2007hd,Kumar:2009vs,
Cacciapaglia:2011kz,Perelstein:2011ez,AguilarSaavedra:2011ck,Beck:2015cga,
Alvarez:2016nrz,Alvarez:2019uxp,Alvarez:2021hxu,Carpenter:2021vga}. For instance,
in supersymmetric theories the signal is enhanced by cascade decays of
gluino-pair production~\cite{Nilles:1983ge,Farrar:1978xj,Toharia:2005gm,
Alves:2011wf}. Due to the large Yukawa coupling $y_t$ of the top quark, the
production of the $pp\to t\bar{t}t\bar{t}$ process is also of high importance to
study modifications of the Higgs sector. For example, the production of heavy
Higgs bosons in association with top-quark pairs in two-Higgs-doublet
models~\cite{Dicus:1994bm,Craig:2015jba,Craig:2016ygr} can have a big impact on
this production rate. Similar effects have been also found in certain top-philic
Dark Matter models~\cite{Boveia:2016mrp,Albert:2017onk}. Furthermore, in some
composite-Higgs models the top quark is not a fundamental particle and its
compositeness can be studied at lower energies via the impact of
higher-dimensional operators of an effective field theory and to which the $pp\to
t\tb t\tb$ process is particularly
sensitive~\cite{Giudice:2007fh,Pomarol:2008bh,Banelli:2020iau}.
 
Given the strong sensitivity of the four top-quark production process to
modifications of the Standard Model (SM) dynamics, a measurement of this process
can also provide stringent constraints on four-fermion operators if interpreted
within an effective field theory approach~\cite{Zhang:2017mls,
AguilarSaavedra:2018nen,Hartland:2019bjb,Banelli:2020iau,Ethier:2021bye} or
within simplified models~\cite{Darme:2018dvz,Cao:2021qqt}.  
However, the former should be taken with care in the presence of on-shell
intermediate resonances~\cite{Darme:2021gtt}.
Finally, the four top-quark final
state can also be utilized to determine CP properties of the SM Higgs
boson~\cite{Cao:2016wib,Cao:2019ygh}.

In summary, even though the production of four top-quark events is one of the
rarest and most energetic signatures at the LHC, it represents an extremely
versatile process for testing the consistency of the SM and put further
constraints on BSM physics. Nonetheless, top quarks are short-living unstable
particles that decay predominantly into a $W$ boson and a bottom quark giving
rise to a $W^+W^-W^+W^- b\bar{b}b\bar{b}$ final state. After accounting for the
decays of the $W$ bosons the \tttt{} final state yields signatures of
unprecedented complexity with at least $12$ final state particles, which
typically comprise four $b$ jets and additional light jets or multiple leptons.
Therefore, in order to harness the full potential of the $pp\to t\bar{t}t\bar{t}$
production process, reliable theoretical predictions for the SM process are
necessary.

The dominant next-to-leading order (NLO) QCD corrections at
$\mathcal{O}(\alpha_s^5)$ have been calculated in Ref.~\cite{Bevilacqua:2012em}
for the first time and later revised in Ref.~\cite{Maltoni:2015ena}.  The
complete NLO corrections including electroweak (EW) corrections as well as all
subleading contributions to the \tttt{} production process at perturbative orders
from $\mathcal{O}(\alpha^5)$ to $\mathcal{O}(\alpha_s^5)$ have been computed for
the first time in Ref.~\cite{Frederix:2017wme}. Furthermore, a comparison of
parton shower matched computations in the MC@NLO~\cite{Frixione:2002ik,
Frixione:2003ei} framework as provided by \mg5{}~\cite{Alwall:2014hca,Frederix:2018nkq} and
\linebreak{} \sherpa{}~\cite{Gleisberg:2008ta,Sherpa:2019gpd} has been presented
in Ref.~\cite{ATLAS:2020esn}.

In this paper we present a state-of-the-art Monte Carlo event generator
implemented in the \powhegbox{} V2 framework~\cite{Nason:2004rx,Alioli:2010xd}. We
consider next-to-leading QCD corrections for the $pp \to t\bar{t}t\bar{t}$
production at $\mathcal{O}(\alpha_s^5)$, while also including all subleading EW
production modes at leading-order (LO) accuracy. Furthermore, we model top-quark
decays at LO while retaining spin-correlation effects. 
Having multiple event generators at hand allows to study in more detail various
sources of theoretical uncertainties intrinsic to the different approaches of
matching fixed-order NLO QCD calculations to parton showers.  Our generator is
publicly available on the \powhegbox{}
website\footnote{\url{https://powhegbox.mib.infn.it}}.

The paper is organized as follows. In section~\ref{sec:nlocomput} we provide a
brief overview of the technical aspects of this work together with a short review
of the structure of higher-order corrections for the four top-quark production at
hadron colliders. Next, we discuss in section~\ref{sec:setup} the numerical setup
for the theoretical predictions shown in section~\ref{sec:results}. At last, we
will give our conclusions in section \ref{sec:conclusions}.

\section{Technical aspects}
\label{sec:nlocomput}
In this section we introduce our new event generator and elaborate on the
technical aspects of its implementation in the \powhegbox{} framework. Before
doing so, we review the general structure and importance of the various
higher-order corrections at the one-loop level. 

\subsection{Anatomy of higher-order corrections to $pp \to t\bar{t}t\bar{t}$}
\begin{figure*}[ht!]
 \centering
 \includegraphics[height=4cm]{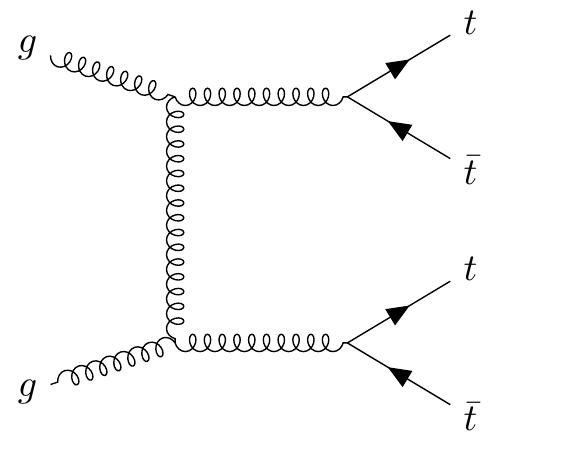}
 \includegraphics[height=4cm]{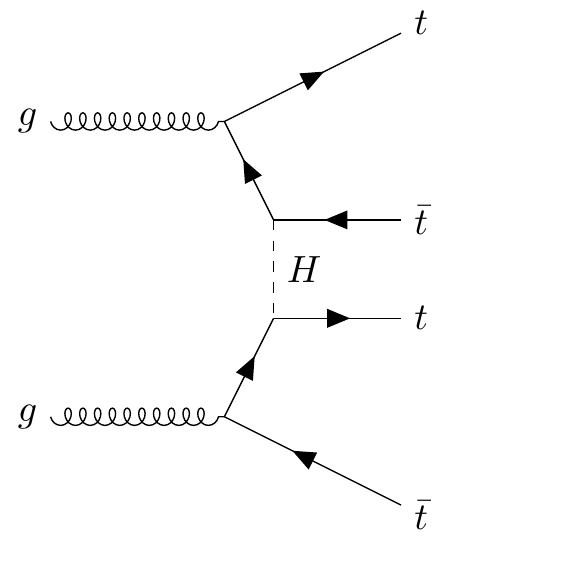}
 \includegraphics[height=4cm]{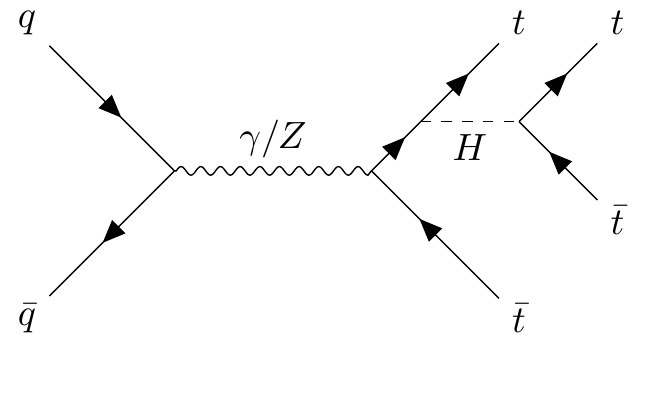}
 \caption{Representative Feynman diagrams for the tree-level amplitudes at 
 $\mathcal{O}(\alpha_s^2)$ (left), $\mathcal{O}(\alpha_s\alpha)$ (middle), 
 $\mathcal{O}(\alpha^2)$ (right).}
 \label{fig:fd}
\end{figure*}

We start by reviewing the findings of Ref.~\cite{Frederix:2017wme} to motivate
our choice to also include a subset of the most important subleading
contributions. At tree-level the $pp\to t\tb t\tb$ process receives contributions
at various orders of the strong $(\alpha_s)$ and electromagnetic $(\alpha)$
coupling. Therefore, the perturbative expansion of the \linebreak{} leading-order
cross section takes the form
\begin{equation}
\begin{split}
 \sigma_{t\bar{t}t\bar{t}}^\lo =~&\alpha_s^4\Sigma_{4,0} + 
 \alpha_s^3\alpha~\Sigma_{3,1}  \\
 +~&\alpha_s^2\alpha^2\Sigma_{2,2} + \alpha_s\alpha^3~\Sigma_{1,3} 
 + \alpha^4\Sigma_{0,4} \\
 =~&\Sigma^\lo_1 + \Sigma^\lo_2 + \Sigma^\lo_3 + \Sigma^\lo_4 + \Sigma^\lo_5\;,
\label{eqn:sigLO}
\end{split}
\end{equation}
where $\Sigma^\lo_1$, $\Sigma^\lo_3$ and $\Sigma^\lo_5$ correspond to squared
amplitude contributions and are therefore strictly positive. In
Fig.~\ref{fig:fd} we depict a few sample Feynman diagrams for these
contributions. On the contrary, the terms $\Sigma^\lo_2$ and $\Sigma^\lo_4$
originate mostly from the interference of Feynman diagrams from different
perturbative orders, for the $gg$ and $q\bar{q}$ initial states, and thus are
not necessarily positive. Furthermore, they receive contributions from photon
initiated processes.

An atypical feature of the $pp\to t\tb t\tb$ process is the presence of both the
$q\bar{q}$ and the $gg$ initiated production channels in the EW contributions up
to $\mathcal{O}(\alpha_s^2\alpha^2)$. The subleading terms at
$\mathcal{O}(\alpha_s\alpha^3)$ and $\mathcal{O}(\alpha^4)$ contribute only to
$q\bar{q}$ channels.
The size of these EW contributions has been studied in
Ref.~\cite{Frederix:2017wme}, where it has been found that the $\Sigma^\lo_2$ and
$\Sigma^\lo_3$ contributions can be as large as $-30\%$ and $+40\%$ respectively
relative to the leading term, $\Sigma^\lo_1$. The large contributions stem from
the $tt \to tt$ scattering, for which a representative Feynman diagram is shown
in the middle panel of Fig.~\ref{fig:fd}. The exchange of massive bosons, such as
the Higgs or $Z$ boson, between non-relativistic top quarks gives rise to
Sommerfeld enhancements, which have been thoroughly studied for top-quark pair
production in Refs.~\cite{Kuhn:2013zoa,Beneke:2015lwa}. Contrarily, the terms
$\Sigma^\lo_4$ and $\Sigma^\lo_5$ only give rise to corrections below $1\%$. Even
though, some of these, formally subleading, contributions can be very large the
total correction  amounts to roughly positive $5\%-15\%$ depending on the chosen
renormalization scale as there are strong cancellations among them. 

At the next-to-leading order the situation becomes rather complex and the
perturbative expansion of the cross section ranges from $\mathcal{O}(\alpha^5)$
to $\mathcal{O}(\alpha_s^5)$:
\begin{multline}
 \sigma_{t\bar{t}t\bar{t}}^\nlo = \alpha_s^5\Sigma_{5,0} + 
 \alpha_s^4\alpha~\Sigma_{4,1} + \alpha_s^3\alpha^2\Sigma_{3,2}  \\
 + \alpha_s^2\alpha^3\Sigma_{2,3} + \alpha_s\alpha^4\Sigma_{1,4} 
 + \alpha^5\Sigma_{0,5}\;.
\end{multline}
As all LO production channels are non-vanishing the classification into pure QCD
and pure EW corrections breaks down. The complexity of the computation of NLO
corrections in this case increases tremendously because QCD and QED infrared
singularities have to be subtracted simultaneously. Only the
$\mathcal{O}(\alpha_s^5)$ contribution can be considered a pure QCD correction
and represents the bulk of the NLO corrections. The sum of all subleading NLO
corrections is below $5\%$ at the inclusive level. However, partial contributions
can be sizable as they compensate for the large scale dependence at the leading
order. At the differential level NLO corrections to EW channels can become
sizable nonetheless, as demonstrated in Ref.~\cite{Frederix:2017wme},
especially in the thresold region. In this region, however, we expect the
results based on a fixed-order calculation to be less reliable due to the
appearance of large logarithms that ultimately need to be resummed.  At future
colliders operating at higher energies the subleading EW corrections contribute
at the same level to inclusive cross sections, while their relative
contributions for differential observables can be much more important if these
are sensitive to the threshold region.

In summary, besides the NLO QCD corrections at $\mathcal{O}(\alpha_s^5)$, the
dominant contribution for the production of four top quarks are the subleading
production channels at LO that amount to roughly a $+10\%$ correction followed
by the remaining NLO corrections. The theoretical uncertainties due to the
renormalization and factorization scale dependence are also dominated by the
leading NLO QCD corrections. In this work we consider NLO QCD corrections and
the subleading EW production channels but not the higher-order corrections to
those channels. Furthermore, we neglect photon initiated processes as their
contributions are suppressed by the photon parton distribution function. In
view of the complexity and size of NLO EW corrections we deem them as
dispensable for now. Their inclusion will be beneficial at a later stage as
they are expected to reduce the theoretical uncertainties of the subleading
contributions here included at LO accuracy.

\subsection{Implementation and validation}
We proceed with the implementation details and cross checks that we performed to
validate our calculation.

The \powhegbox{} framework already provides all process independent ingredients
for computing NLO QCD corrections. The necessary tree-level and one-loop matrix
elements at $\mathcal{O}(\alpha_s^4)$ and $\mathcal{O}(\alpha_s^5)$ are taken
from \openloops{}~\cite{Cascioli:2011va,Buccioni:2017yxi,Buccioni:2019sur}, which
also provides the necessary spin- and color-correlated born matrix elements to
construct the infrared subtraction terms. We are using the \openloops{} interface
to \powhegbox{} introduced in Ref.~\cite{Jezo:2016ujg}. The matrix elements for
the subleading leading-order electroweak contributions for the $gg \to t\tb
t\tb$ and $q\qb \to t\tb t\tb$ processes are instead extracted from \mg5{}.
Technically, they are included as part of the virtual corrections and,
therefore, do not affect the generation of radiation in the \powhegbox{}
framework. Neither are they considered in the generation of the color
assignment of the underlying born event. This is expected to have a very
limited impact, because the EW contributions do not introduce any new basis
elements in the colorflow decomposition of the tree-level matrix elements.
Thus, our approach is an approximation that treats the subleading EW
contributions as an inclusive correction applied to the leading QCD matrix
elements. We have confirmed the validity of this approximation by an explicit
leading-order parton-shower matched computation using full tree-level matrix
elements but assigned colorflows either according to the QCD or the full matrix
elements. We found very good agreement between these two options with
differences smaller than $2\%$ at the differential level.

The decays of the top-quarks are included using the algorithm presented in
Ref.~\cite{Frixione:2007zp}, which retains spin correlations at leading-order
accuracy. On the technical side, the implementation is a straightforward
adaptation of the algorithm used in Ref.~\cite{Cordero:2021iau}, which also
allows to include off-shell virtualities for resonant particles. For more
details on the decay implementation, we refer the reader to the aforementioned
references. All necessary QCD decay-chain matrix elements are also taken from
\mg5{}. Again, we have checked explicitly at leading-order that the structure
of the EW contributions, does not modify the spin-correlations effects in the
top decay. 

We have validated our implementation by performing several cross checks. For
instance, we compared all tree-level and one-loop matrix elements against
\linebreak{} \mg5{} at a few phase space points. At fixed-order we reproduced the
inclusive cross sections at NLO QCD as given in Ref.~\cite{Bevilacqua:2012em} and
Ref.~\cite{Frederix:2017wme}. The integrated leading-order cross section
including all electroweak production channels, as given by Eq.~\eqref{eqn:sigLO},
has been checked against \phegas{}
\cite{Kanaki:2000ey,Papadopoulos:2000tt,Cafarella:2007pc}. And finally, the
algorithm for the decay of the four top final state has been validated at the
differential level against \mg5{} in conjunction with \madspin{}
\cite{Artoisenet:2012st}. To be precise, we compared leptonic observables in
the fully leptonic decay channel at LO accuracy, ignoring the EW tree-level
contributions for a moment. We find excellent agreement in all cases once spin
correlations have been taking into account.

\section{Computational setup}
\label{sec:setup}
We consider the production of four top-quarks at the LHC with a center-of-mass
energy of $\sqrt{s} = 13$ TeV. For this study we fix the SM parameters to the
following values
\begin{equation}
\begin{array}{ll}
  G_F = 1.166378\cdot 10^{-5}~\GeV^{-2}\;, &  M_t = 172.5~\GeV\;, \\[0.1cm]
 M_W = 80.385~\GeV\;, &  \Gamma_W = 2.09767~\GeV\;, \\[0.1cm] 
 M_Z = 91.1876~\GeV\;, &  \Gamma_Z = 2.50775~\GeV\;, \\[0.1cm] 
 M_H = 125~\GeV\;, & \Gamma_H = 0.00407~\GeV\;.
\end{array}
\end{equation}
The electromagnetic coupling is derived from the input parameters in the
$G_\mu$-scheme~\cite{Denner:2000bj} and given by
\begin{equation}
 \alpha = \frac{\sqrt{2}}{\pi} G_F M_W^2\left( 1- \frac{M_W^2}{M_Z^2}\right)\;.
\end{equation}
We calculate the top-quark width at NLO accuracy from all the other input
parameters by computing the three-body decay widths $\Gamma(t\to f\bar{f}^\prime
b)$ into any light fermion-pair $f\bar{f}^\prime$ and a massive $b$ quark. To
this end, we employ a numerical routine of the MCFM implementation of
Ref.~\cite{Campbell:2012uf}.
If not explicitly mentioned otherwise all presented results have been obtained
for the NNPDF3.1~\cite{Ball:2017nwa}\footnote{The LHAPDF ID numbers for the PDF
sets are $315200$ at LO and $303400$ at NLO} parton distribution function (PDF)
as provided through the \lhapdf{} interface~\cite{Buckley:2014ana}. We adopt the
dynamical scale choice of Ref.~\cite{Maltoni:2015ena} and choose for the
renormalization and factorization scale
\begin{equation}
 \centering
 \mu_R = \mu_F = \mu_0 = \frac{H_T}{4}\;,
\end{equation}
where
\begin{equation}
 H_T = \sum_{i\in \{t,\tb,t,\tb,j\}} \sqrt{m_i^2+p^2_{T,i}}\;.
 \label{eqn:HT}
\end{equation}
In order to estimate the theoretical uncertainty due to our particular choice of
renormalization and factorization scales, we vary them independently in the range
of 
\begin{multline}
 \left(\frac{\mu_R}{\mu_0}, \frac{\mu_F}{\mu_0}\right) = \Big\{
 (0.5,0.5), (0.5,1), (1,0.5), \\
(1,1), (1,2), (2,1), (2,2)\Big\}\;,
\end{multline}
and take the envelope as uncertainty estimate. Let us note, that we vary these
scales only in the calculation of the hard matrix elements, thus neither the
generation of the hardest emission nor the consecutive parton shower evolution
are directly affected by these variations.

The matching in the \powhegbox{} framework depends on the two damping parameters
$h_\textrm{damp}$ and $h_\textrm{bornzero}$ that split the real matrix elements
into a finite and an infrared singular contribution. While the singular piece is
used to resum soft and collinear QCD splittings the finite part, containing only
hard emissions, is treated at fixed-order. For more details on these
parameters we refer the reader to Refs.~\cite{Alioli:2008gx,Alioli:2008tz,
Alioli:2010xd,Jezo:2018yaf}. Based on experience drawn from previous
work~\cite{Jezo:2018yaf,Cordero:2021iau,Bevilacqua:2021tzp} these parameters are
chosen as
\begin{equation}
 h_\textrm{damp} = \frac{H_T}{4}\;, \qquad h_\textrm{bornzero} = 5\;,
\end{equation}
and $h_\textrm{damp}$ is evaluated on the underlying Born kinematics.
We study the impact of our choice by considering the envelope of the
following independent variations of these parameters
\begin{multline}
 \left( h_\mathrm{damp}, h_\mathrm{bornzero}\right) = \Bigg\{
 \left(\frac{H_T}{4},5\right), \left(\frac{H_T}{4},2\right), \\
 \left(\frac{H_T}{4},10\right), \left(\frac{H_T}{8},5\right),
 \left( \frac{H_T}{2}, 5\right) \Bigg\}\;.
 \label{eqn:std_damp}
\end{multline}

As part of the validation of our computation, we perform a comparison of our
results with those obtained with the \mg5{} framework that employs the MC@NLO
matching to parton showers. We use the same input parameters and renormalization
and factorization scales as discussed above. Furthermore, the matching in the
MC@NLO scheme depends crucially on the choice of the initial shower scale
$\mu_Q$. Here we keep the \mg5{} default choice of
\begin{equation}
 \mu_Q = \frac{H_T}{2}\;.
\end{equation}
We study the dependence on this scale by varying it by a factor of $2$ up and
down.

Finally, for all of the following theoretical predictions we use
\pythia{}~\cite{Sjostrand:2006za,Sjostrand:2014zea} (v.8.306) to perform the
shower evolution. However, effects from matrix element corrections to the decays,
hadronization and multiple interactions are not addressed in this work. The
showered events are analyzed using the
\textsc{Rivet}~\cite{Buckley:2010ar,Bierlich:2019rhm} framework.

\section{Phenomenological results}
\label{sec:results}
In this section we present our theoretical predictions. We start by investigating
the different sources of theoretical uncertainties as well as the impact of
subleading EW contributions for the four top-quark production at the inclusive
level, i.e. for stable top quarks. Afterwards, we present a sample study for
decayed tops in the single-lepton decay channel with a particular focus on the
impact of spin correlations contributions at the differential level.

\subsection{Total cross sections}
We start our discussion with a detailed investigation of the uncertainty budget
of inclusive cross sections. In Table~\ref{tab:xsecs} we show the integrated
cross sections at LO accuracy for the leading QCD contribution in the first
column and the same also including subleading EW channels in the second column.
In the fourth column predictions where NLO QCD corrections on top of the full LO
contributions are taken into account are shown. Besides our default choice
(NNPDF3.1), we also report integrated cross sections for the MMHT
2014~\cite{Harland-Lang:2014zoa} and CT18~\cite{Hou:2019efy} PDF sets. In
addition, we also provide the corresponding theoretical uncertainties from scale
variations, denoted with $\delta_\textrm{scale}$ and internal PDF uncertainties
denoted with $\delta_\textrm{PDF}$.
\begin{table*}[ht!]
\begin{center}
\begin{tabular}{ccccccccc}
  \hline
  &&&&&&&& \\
  PDF & $\sigma^\lo_\textrm{QCD}$ [fb] & $\sigma^\lo$ [fb] & $\delta_\textrm{scale}$ & $\sigma^\nlo$ [fb] & $\delta_\textrm{scale}$ & $\delta_\textrm{PDF}$ & $\mathcal{K} = \frac{\nlo}{\lo}$ \\
  &&&&&&&& \\
  \hline
  NNPDF3.1 & $8.31$  & $8.79$  & $^{+6.07~(69\%)}_{-3.30~(38\%)}$ & $11.65$ & $^{+1.98~(17\%)}_{-2.57~(22\%)}$ & $^{+0.28~(2\%)}_{-0.28~(2\%)}$ & $1.33$ \\
  MMHT     & $10.69$ & $11.19$ & $^{+8.23~(74\%)}_{-4.36~(39\%)}$ & $11.62$ & $^{+1.95~(17\%)}_{-2.54~(22\%)}$ & $^{+0.63~(5\%)}_{-0.53~(5\%)}$ & $1.04$ \\
  CT18     & $10.04$ & $11.04$ & $^{+7.68~(70\%)}_{-4.18~(38\%)}$ & $11.74$ & $^{+1.97~(17\%)}_{-2.56~(22\%)}$ & $^{+0.46~(4\%)}_{-0.36~(3\%)}$ & $1.06$ \\
  \hline
\end{tabular}
\caption{Total cross sections for the $pp \to t\tb t\tb$ process at $\sqrt{s}=13$ 
TeV for the LHC. Cross sections at LO and NLO for various PDF sets are shown 
together with theoretical uncertainties estimates from scale variations and 
internal PDF uncertainties.}
\label{tab:xsecs}
\end{center}
\end{table*}
The size of the subleading EW contributions ranges from $+5\%$ to $+10\%$
depending on the PDF set employed, where for MMHT they have the smallest impact
and for CT18 the largest. At leading order the scale uncertainties are large and
of the order of $70\%$, which easily accounts for the differences between
predictions based on different PDF sets. Including NLO QCD corrections reduces
the scale uncertainties by more than a factor of $2$ to $22\%$. 
The $\mathcal{K}$-factor strongly depends on the PDF set employed at the leading
order. For instance, if LO PDF sets are used we find $+33\%$ corrections in the
case of the NNPDF PDF set and only between $+4\%-6\%$ corrections for MMHT and
CT18. However, when NLO PDF sets are used for both, LO and NLO predictions, we
find a rather large $\mathcal{K}$-factor equal to $1.54$ stable with respect to
the choice of PDF set. Note that the latter $\mathcal{K}$-factor enters the
calculation of the Sudakov form factor of the hardest emission. In view of its
large size the necessity of including higher order corrections in fixed order as
well as parton-shower matched predictions becomes apparent.
The PDF uncertainties are much smaller than the scale uncertainties and have been
estimated to be $\pm 2\%$ up to $\pm 5\%$ depending on the chosen PDF set. Let us
note here, that we follow Ref.~\cite{Bevilacqua:2016jfk} and rescale the PDF
uncertainties of the CT18 PDF set, which are provided at the $90\%$ confidence
level (CL), by a factor of $1/1.645$ in order to make them comparable with the
other PDF sets that provide uncertainties only at the $68\%$ CL. Finally, we also
observe that once NLO QCD corrections are taken into account differences for the
central prediction for each PDF are at the $1\%$ level, while there are sizable
differences at LO.

\subsection{Inclusive differential distributions}
Let us now turn to the discussion of differential distributions at the fully
inclusive level, i.e. for stable top quarks. We do not impose any selection cuts
on the final state top quarks. Jets are defined via the anti-$k_T$ jet
algorithm~\cite{Cacciari:2008gp} with $R=0.4$ as provided by
\textsc{FastJet}~\cite{Cacciari:2005hq,Cacciari:2011ma} and subject to the
following cuts
\begin{equation}
 p_T(j) > 25~\GeV\;, \qquad |y(j)| < 2.5\;.
\end{equation}
Furthermore, we order top quarks, irrespective of them being particle or
anti-particle, according to their transverse momenta in decreasing order.

Differential cross section distributions are shown in the following as plots
containing three panels. The upper panel shows theoretical predictions at NLO
accuracy, the middle panel depicts scale uncertainties computed from an envelope
of independent variations of renormalization and factorization scales, and the
bottom panel illustrates matching uncertainties estimated by varying the various
\powhegbox{} specific damping parameters or the initial shower scale $\mu_Q$ in
the case of the MC@NLO matching scheme in \mg5{}.
All theoretical predictions including NLO QCD corrections are labeled with
\textit{QCD}, while we label predictions that include EW Born contributions as
\textit{QCD+EW}. Furthermore, we do not show theoretical uncertainties for the
QCD-only predictions obtained with the \powhegbox{} as we find that they are very
similar to those when EW contributions are taken into account. Note that even
though the subleading EW channels have separately a sizable scale dependence due
to $\alpha_s^n(\mu_R)$ with $n\leq 3$, the sum of all EW production modes has a
reduced dependence and is smaller than the corresponding scale uncertainty of the
dominant NLO QCD corrections at $\mathcal{O}(\alpha_s^5)$.

%
\begin{figure*}[ht!]
 \centering
 \includegraphics[width=0.495\textwidth]{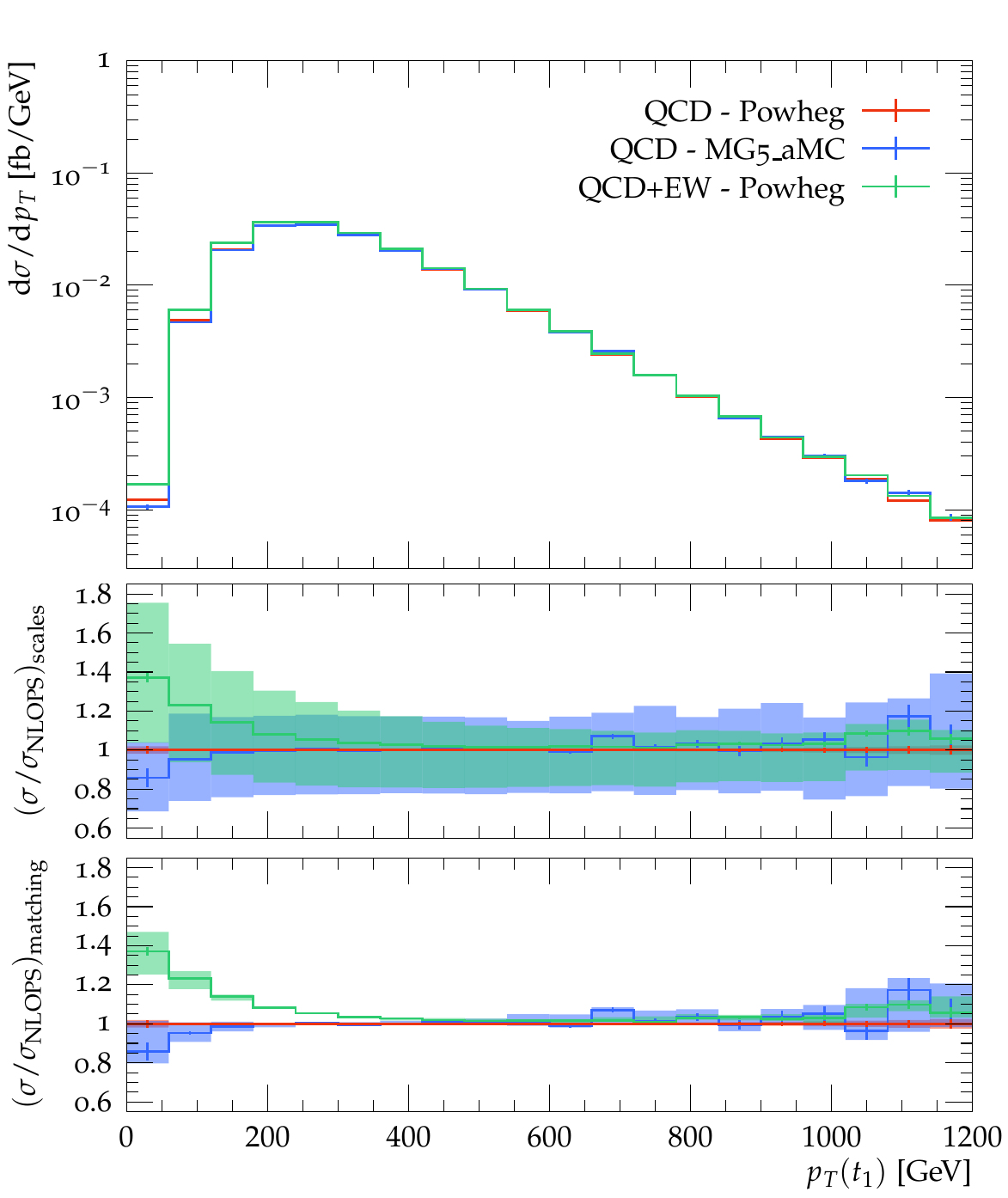}
 \includegraphics[width=0.495\textwidth]{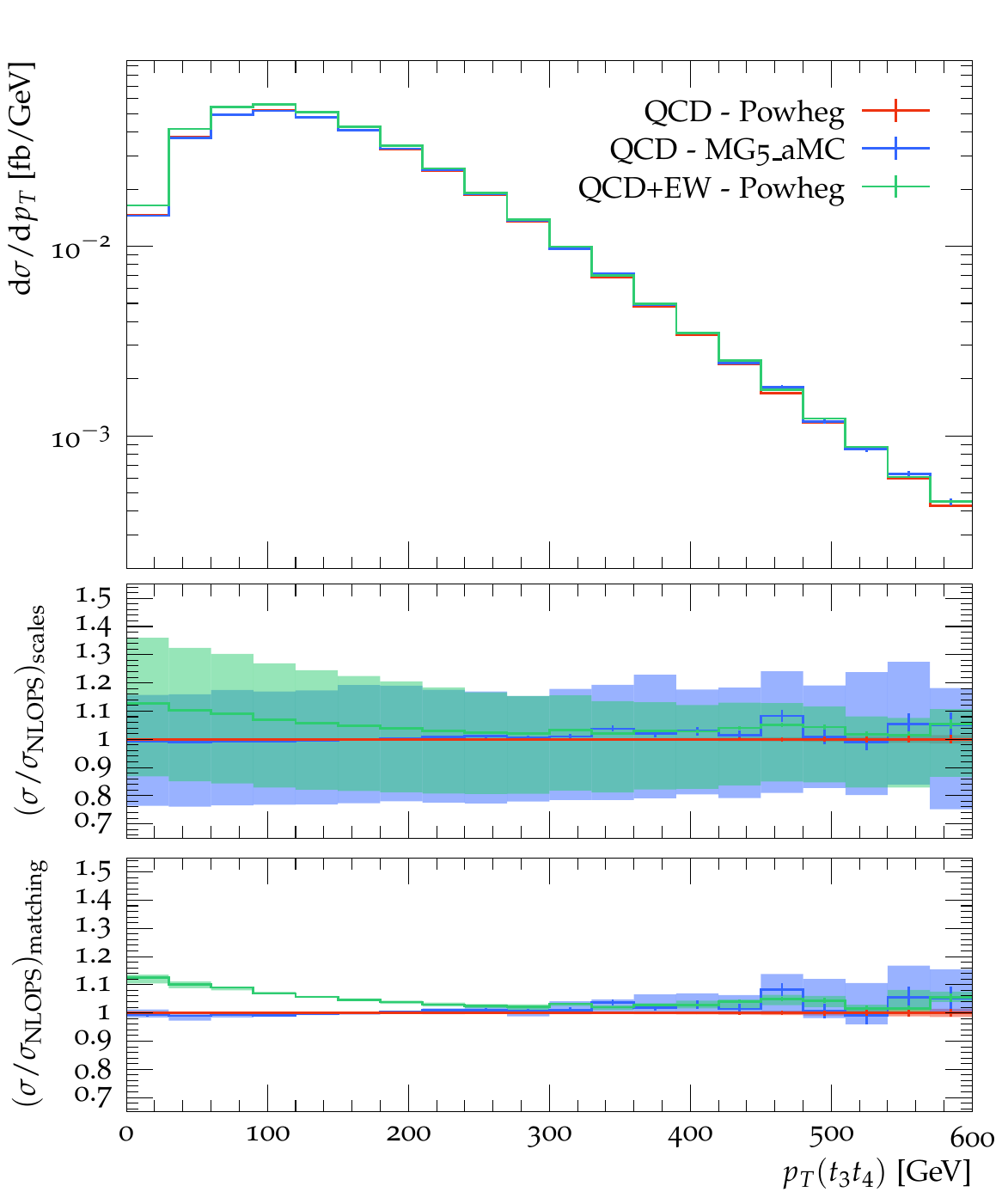}
 \caption{Differential cross section distribution as a function of the transverse
 momentum of the hardest top quark (l.h.s.) and of the third and fourth
 hardest top-quark pair (r.h.s.) for the $pp\to t\tb t\tb$ 
 process.The uncertainty bands correspond to independent variations of the 
 renormalization and factorization scales (middle panel) and of the matching 
 parameters (bottom panel).}
 \label{fig:inc_1}
\end{figure*}
For the transverse momentum of the hardest top quark shown on the left of
Fig.~\ref{fig:inc_1}, we find very good agreement between \mg5{} and the QCD only
prediction of \powhegbox{}. Only in the threshold region, where the transverse
momenta of all four top quarks become small, differences at the level of $15\%$
are visible. However, this phase space region is also subject to sizable matching
uncertainties of the order of $10\%$ for all predictions. In addition, we observe
that the EW LO processes give rise to sizable corrections in the threshold region
of up to nearly $+40\%$. However, for transverse momenta larger than $300$ GeV
the EW contributions modify the spectrum by less than $3\%$. For all theoretical
predictions the scale uncertainties due to missing higher-order corrections are
the dominant source of uncertainty over the whole range of the distribution. In
the case of \mg5{} they range from $\pm 20\%$ in the threshold region to $\pm
30\%$ in the tail of the distribution. In \powhegbox{}, the uncertainties are
slightly smaller and their pattern is inverted ranging from $\pm 28\%$ at the
beginning of the plotted spectrum to $\pm 15\%$ at the end. We checked explicitly
that the increased scale dependence in the threshold region is not due to the
inclusion of the EW contributions at leading order accuracy.  Therefore, the
differences should be rather associated to the different matching frameworks.

Let us now turn to the transverse momentum of the combined third and fourth
hardest top quark depicted in the right panel of Fig.~\ref{fig:inc_1}. Here we
find a remarkable agreement between \mg5{} and \powhegbox{} over the whole
plotted range if only QCD contributions are taken into account. However, also
here we observe sizable corrections of $+12\%$ from the EW contributions for
small values of the transverse momentum while they are at most of the order of
$+3\%$ above $p_T \approx 250$ GeV. As in the previous case, the uncertainties
due to missing higher-order corrections dominate over those due to matching for
most of the plotted spectrum. For the \powhegbox{} prediction the scale
uncertainties start out very symmetric with $\pm 20\%$ but grow more and more
asymmetric towards the end of the plotted range with estimated uncertainties of
$+5\%$ and $-20\%$. A similar trend is not visible for \mg5{} predictions, as
uncertainties are slightly asymmetric from the start with $-25\%$ and $+16\%$ for
small transverse momenta and $-30\%$ and $+12\%$ uncertainties in the tail of the
distribution. For \powhegbox{} matching uncertainties never exceed $\pm 3\%$,
while for \mg5{} they become as large as $10\%$ in the tail of the distribution.

%
\begin{figure*}[ht!]
 \centering
 \includegraphics[width=0.495\textwidth]{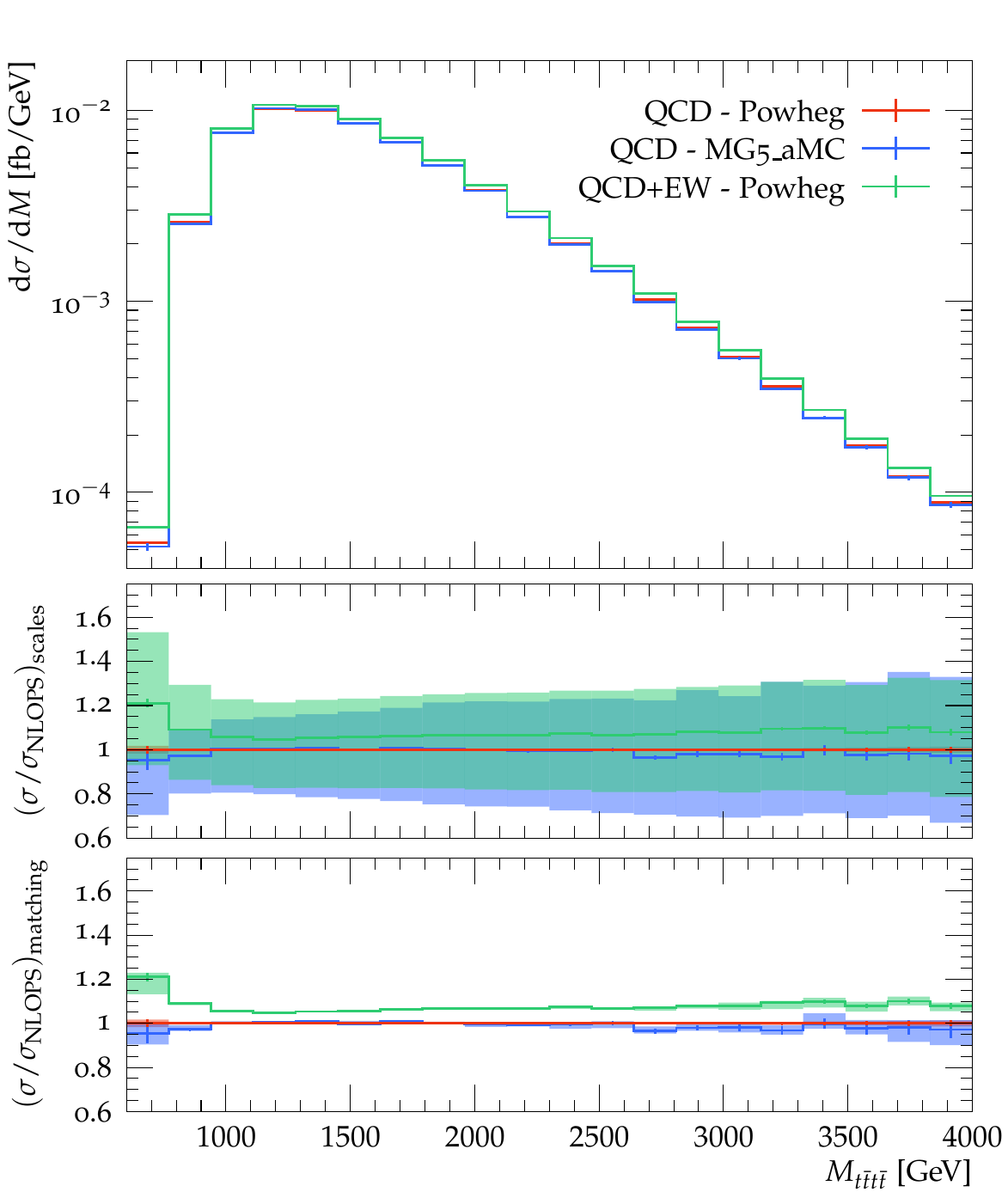}
 \includegraphics[width=0.495\textwidth]{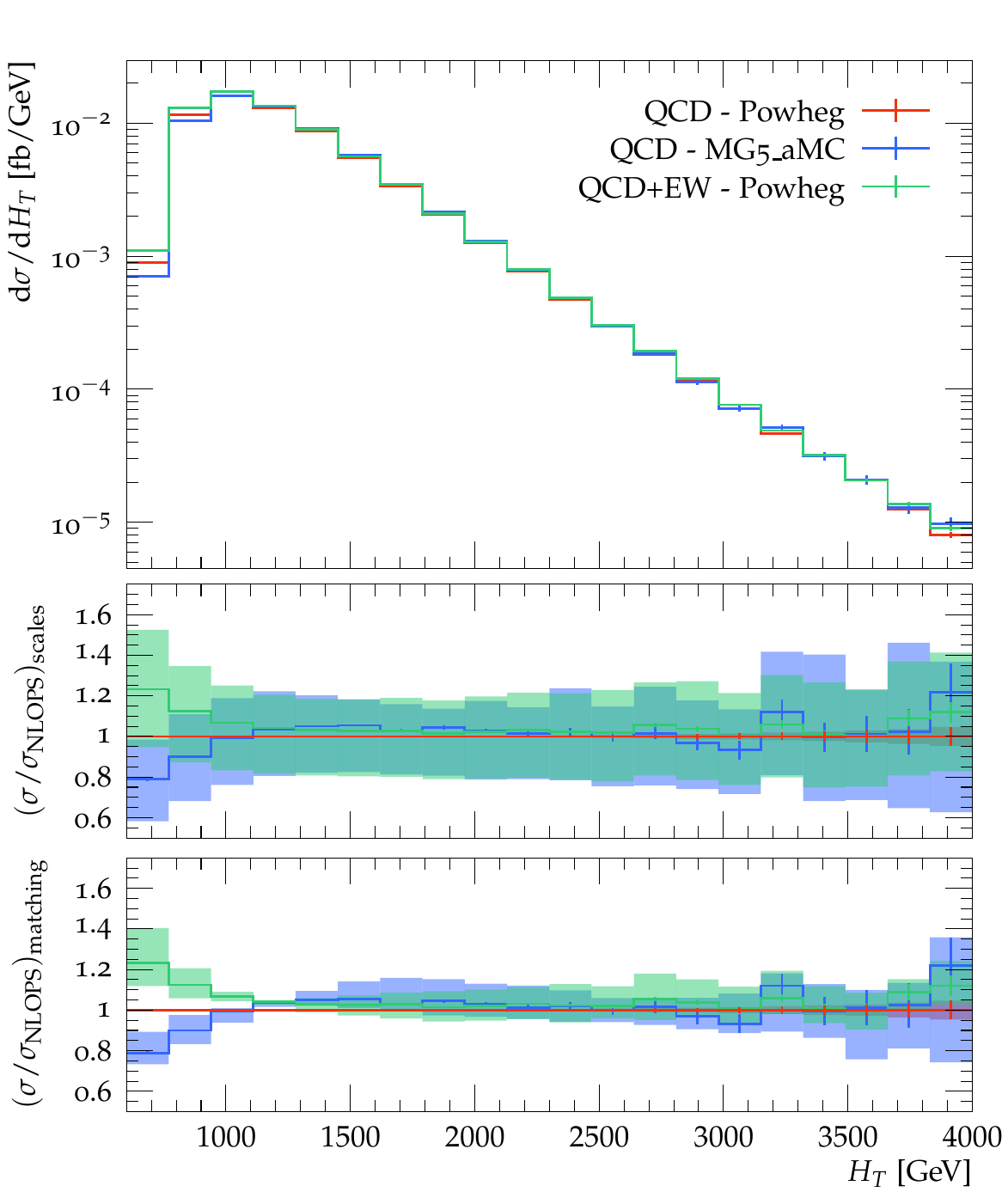}
 \caption{Differential cross section distribution as a function of the invariant
 mass of the four top quarks (l.h.s.) and of the $H_T$ observable (r.h.s.) for the 
 $pp\to t\tb t\tb$ process.The uncertainty bands correspond to independent 
 variations of the renormalization and factorization scales (middle panel) and of
 the matching parameters (bottom panel).}
 \label{fig:inc_2}
\end{figure*}
Next we discuss the invariant mass distribution of the four top quarks shown in
the left panel of Fig.~\ref{fig:inc_2}. We observe that the EW contributions are
sizable not only at the production threshold $M_{t\tb t\tb} \sim 4m_t$ but also
in the high-energy tail of the distribution. To be specific, we find sizable
corrections of the order of $+20\%$ in the threshold region that then decrease
down to $+5\%$ at intermediate value of the invariant mass of around $1.2$ TeV.
However, the leading order EW production channels give rise to $+10\%$
corrections in the very high-energetic tail of the distribution. The matching
uncertainties are for all theoretical predictions small and estimated to be below
$5\%$, where in the case of \powhegbox{} they are the smallest. However, the
uncertainties originating from the choice of renormalization and factorization
scales are between $20\%-30\%$ over the whole spectrum.

We now turn to the $H_T$ distribution, as defined in Eq.~\eqref{eqn:HT}, depicted
in the right panel of Fig.~\ref{fig:inc_2}. Independently of the employed
matching scheme and the inclusion of EW contributions, we find excellent
agreement between all predictions above $H_T \approx 1.2$ TeV with differences
below $3\%$. Only in the first three bins we encounter sizable differences
between the various approaches. For instance, the \mg5{} prediction is up to
$20\%$ smaller at the beginning of spectrum if compared to the corresponding
\powhegbox{} prediction that takes into account the same perturbative
corrections. We observe that EW contributions are sizable here as well and yield
a $+23\%$ enhancement in the low tail. Matching uncertainties are larger than in
the previous observables and reach up to $\pm 10\%$ in the case of \powhegbox{},
and up to $\pm 25\%$ in the case of \mg5{}. On the other hand, scale
uncertainties are at most $\pm 25\%$ for \powhegbox{} predictions and in the case
of \mg5{} they reach $\pm 30\%$. 

%
\begin{figure*}[ht!]
 \centering
 \includegraphics[width=0.495\textwidth]{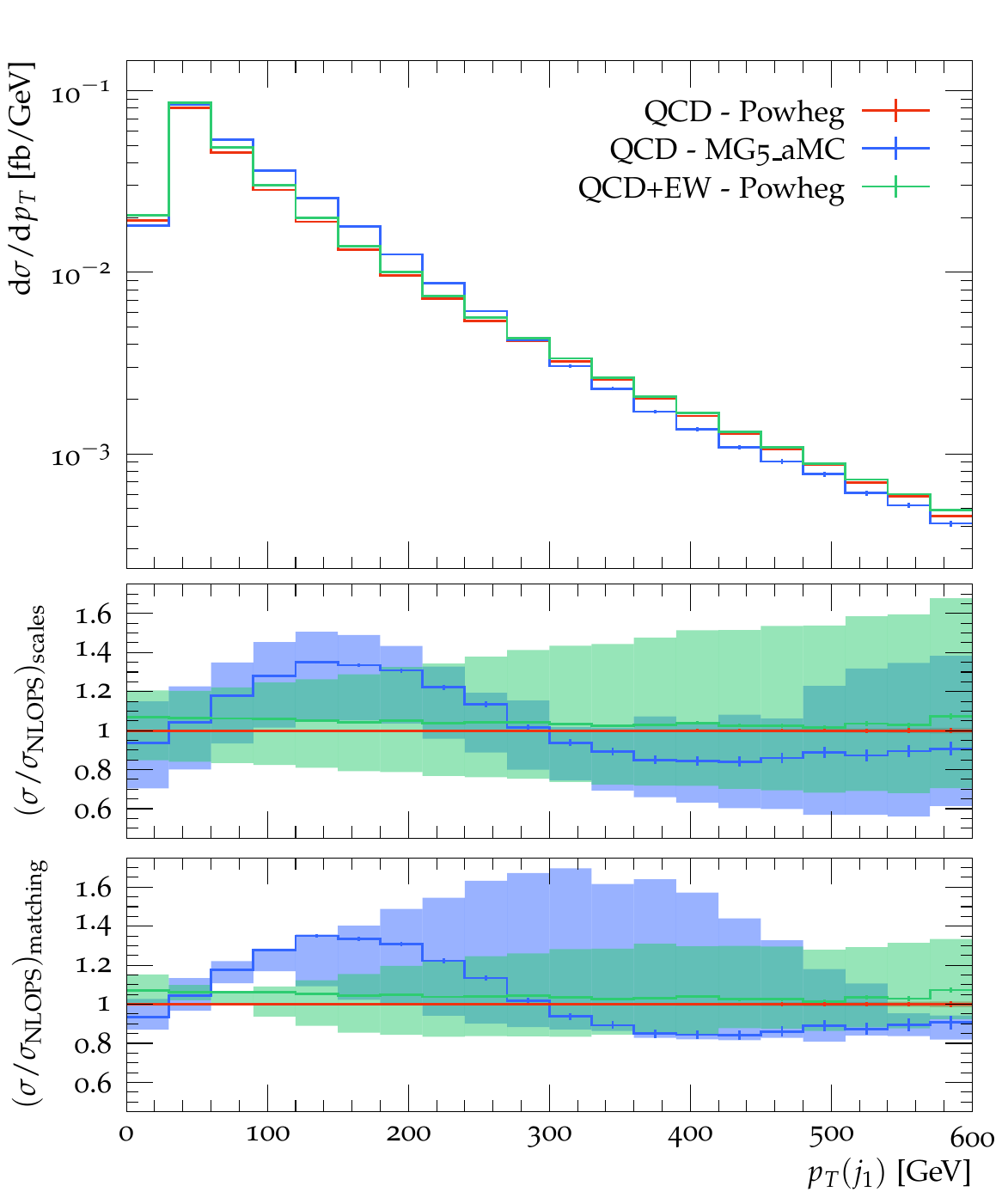}
 \includegraphics[width=0.495\textwidth]{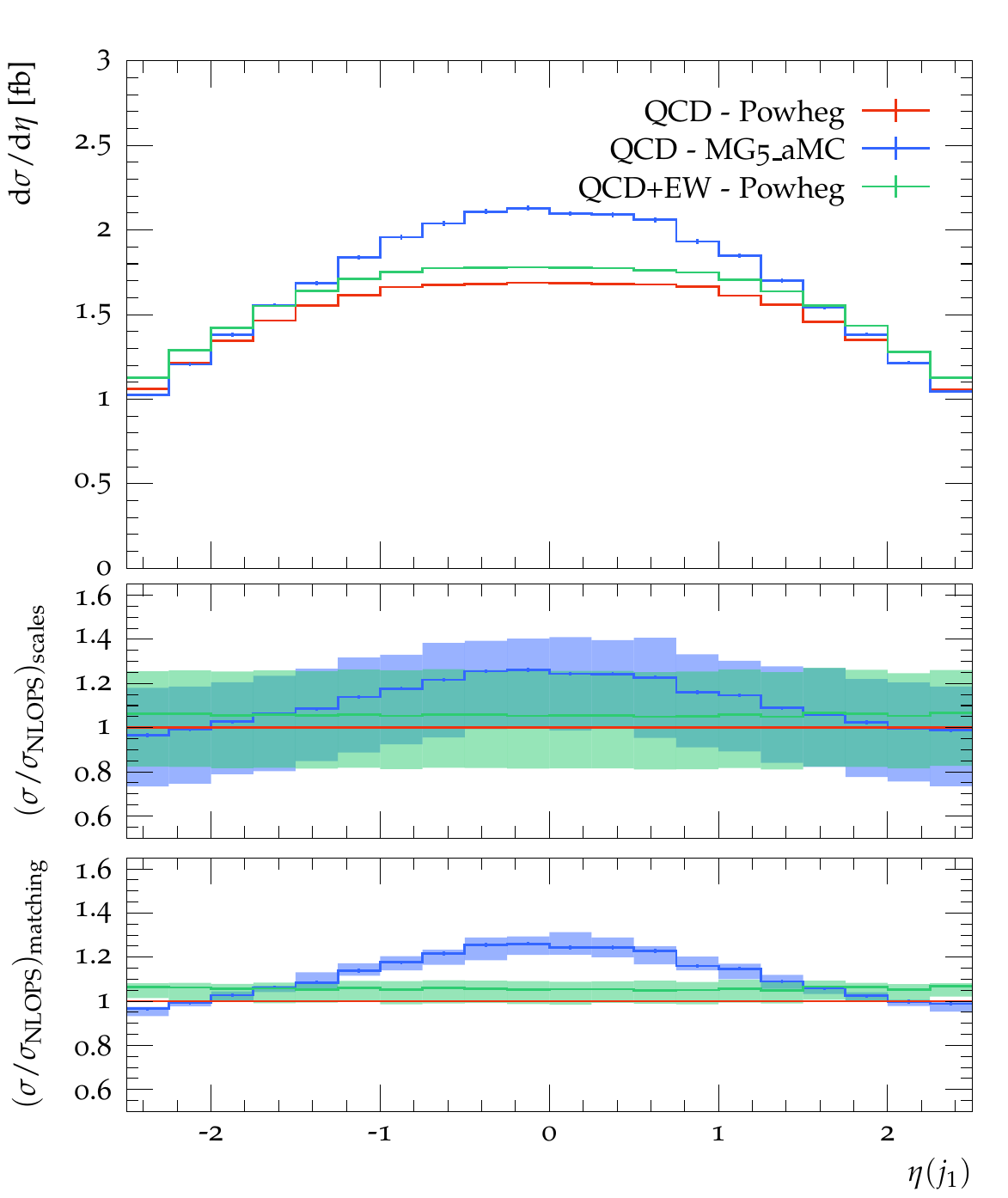}
 \caption{Differential cross section distribution as a function of the transverse
 momentum (l.h.s.) and of the pseudorapidity (r.h.s.) of the hardest jet for the 
 $pp\to t\tb t\tb$ process.The uncertainty bands correspond to independent 
 variations of the renormalization and factorization scales (middle panel) and of
 the matching parameters (bottom panel).}
 \label{fig:inc_3}
\end{figure*}
Finally, we discuss the transverse momentum and the pseudorapidity distribution
of the leading jet as shown in Fig.~\ref{fig:inc_3}. Contrary to all previous
observables, these two distributions are only predicted at leading order
accuracy. Nonetheless, they are useful for exploring differences between the
POWHEG and the MC@NLO matching schemes. Let us start with the transverse momentum
distribution. Here we find large differences between \mg5{} and \powhegbox{} up
to $35\%$ at transverse momenta of the order of $150$ GeV. Furthermore, \mg5{}
predicts a softer spectrum in the tail of the distribution. The corresponding
theoretical uncertainties are large as well and their bands include predictions
from the other event generators, respectively, throughout the majority of the
spectrum. In all cases scale uncertainties start around $\pm 25\%$ and grow up to
$\pm 55\%$ at the end of the plotted range. In addition, the spectrum exhibits a
sizable dependence on the parton-shower matching related parameters. In the case
of \powhegbox{} matching uncertainties are estimated to be of the order of $\pm
6\%$ at the beginning and increase up to $\pm 25\%$ at the end of the
distribution. On the other hand, for \mg5{} the initial shower scale dependence
is rather different, while the beginning and end of the spectrum have modest
uncertainties of the order of $\pm 10\%$ they grow as large as $95\%$ in the
intermediate range. Thus, the choice of initial shower scale $\mu_Q$ has a severe
impact on the shape of the distribution.

At last, let us turn to the pseudorapidity distribution as depicted on the right
of Fig.~\ref{fig:inc_3}. Also for this observable, we find sizable difference
between the various predictions. The impact of the EW production modes is modest
and yields rather constant positive corrections at the level of $5\%-7\%$ over
the whole spectrum. The largest differences are found when different matching
schemes are employed. The \mg5{} predictions is considerably larger by nearly
$25\%$ in the central rapidity region as compared to the \powhegbox{} prediction
that also includes only pure QCD corrections. In all cases, the scale uncertainty
is the dominant contribution to the theoretical uncertainty and amounts to a
constant $\pm 20\%$ over whole plotted range.

Similar differences in the modeling of the leading jet between \mg5{} and the
\powhegbox{} have been already observed for the $pp\to t\tb b \bar{b}$ and $pp\to
t\tb W^\pm$ processes, as discussed in
Refs.~\cite{LHCHiggsCrossSectionWorkingGroup:2016ypw,Buccioni:2019plc,
Cordero:2021iau,Bevilacqua:2021tzp}.

\subsection{Single lepton plus jets signature}
In the following we study a single lepton plus jets signature in order to
investigate the impact of spin-correlated top-quark decays and the impact of the
leading order EW contributions at the fiducial level. The signature is
characterized by the presence of exactly one charged lepton $\ell$, with
$\ell=e,\mu$, at least $4\ b$ jets and at least $6$ light jets. The lepton has to
fulfill $p_T(\ell) > 15$ GeV and $|y(\ell)| < 2.5$. Jets are formed using the
anti-$k_T$ jet algorithm with $R=0.4$ and a jet is labeled a $b$ jet if at least
one of its constituents is a heavy $b$ quark. Light as well as $b$ jets have to
pass the $p_T(j) > 25$ GeV and $|y(j)| < 2.5$ cuts. The definition of the
fiducial phase space volume is inspired by Ref.~\cite{ATLAS:2021kqb}.

We show in the following only theoretical predictions obtained with our
\powhegbox{} implementation. We consider three predictions: one prediction that
includes both spin correlations in the decay of the top quark as well as the
subleading EW channels, and two predictions with either the first or the second
improvement switched off. If spin correlations are omitted the decays of top
quarks and $W$ bosons are generated via independent $1\to 2$ decays.  We do not
discuss matching uncertainties here anymore as we have seen in the previous
section that theoretical uncertainties are dominated by missing higher-order
corrections.  Moreover, matching uncertainties are expected to be very similar
between the various predictions as they are all based on \powhegbox{}.

For the integrated fiducial cross section we obtain for the three approaches the
following results:
\begin{equation}
\begin{split}
 \sigma^\textrm{spin}_\textrm{QCD} &= 0.618^{+0.119~(19\%)}_{-0.142~(23\%)}~\textrm{fb}\;, \\
 \sigma^\textrm{spin}_\textrm{QCD+EW} &= 0.649^{+0.117~(18\%)}_{-0.144~(22\%)}~\textrm{fb}\;, \\
 \sigma^\textrm{no-spin}_\textrm{QCD+EW} &= 0.625^{+0.114~(18\%)}_{-0.139~(22\%)}~\textrm{fb}\;.
\end{split}
\end{equation}
We observe that EW contributions and spin-correlated decays have opposite effects
on the fiducial cross section. For spin-correlated decays the electroweak
production modes increase the cross section by $5\%$ with respect to the QCD only
predictions. On the other hand, if spin correlations are ignored then cross
sections decreases by $4\%$. Even though all predictions for integrated cross
sections are compatible within the scale uncertainty of roughly $\pm 20\%$, we
will see in the following that this is not necessarily the case at the
differential level.

%
\begin{figure*}[ht!]
 \centering
 \includegraphics[width=0.495\textwidth]{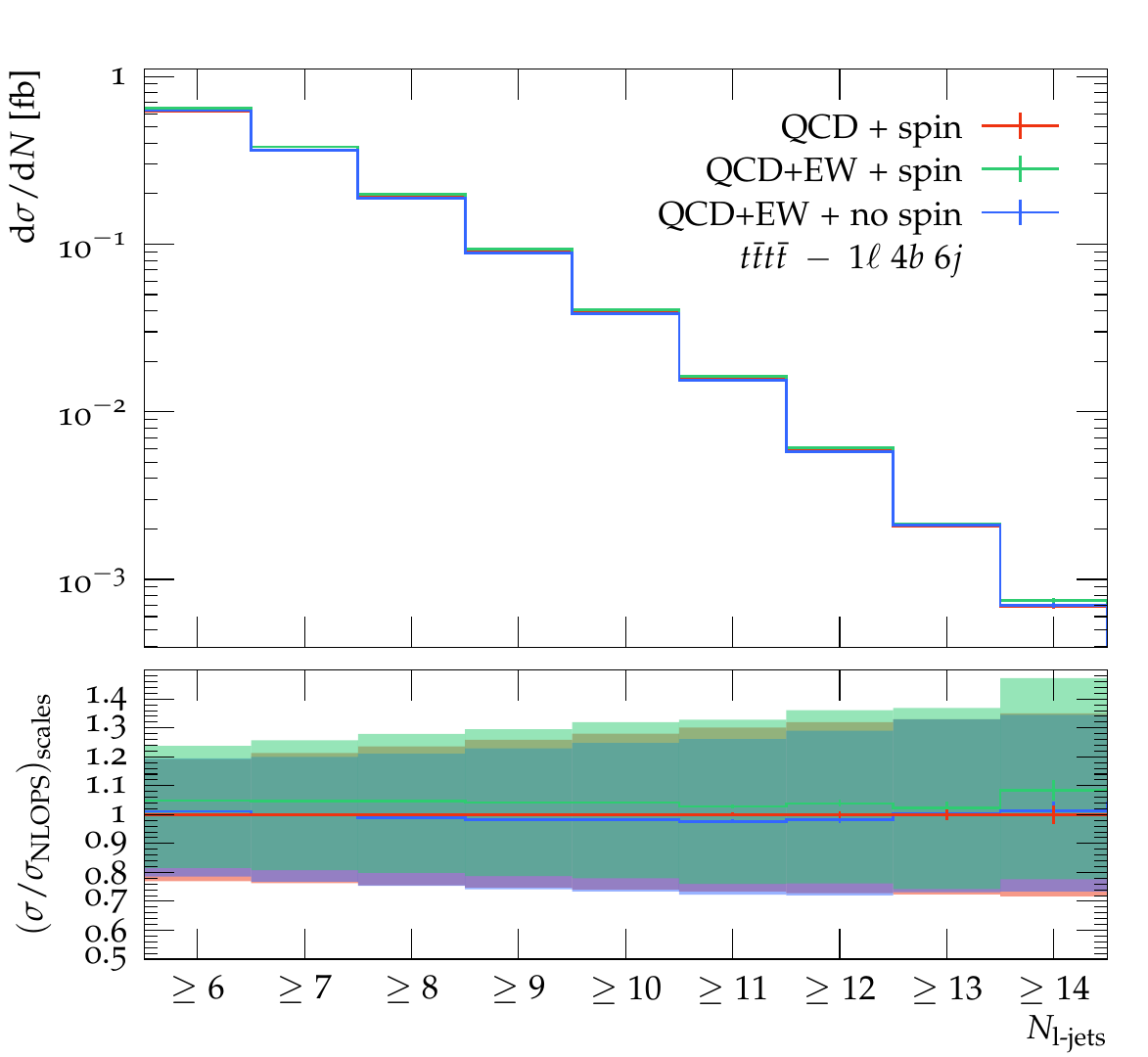}
 \includegraphics[width=0.495\textwidth]{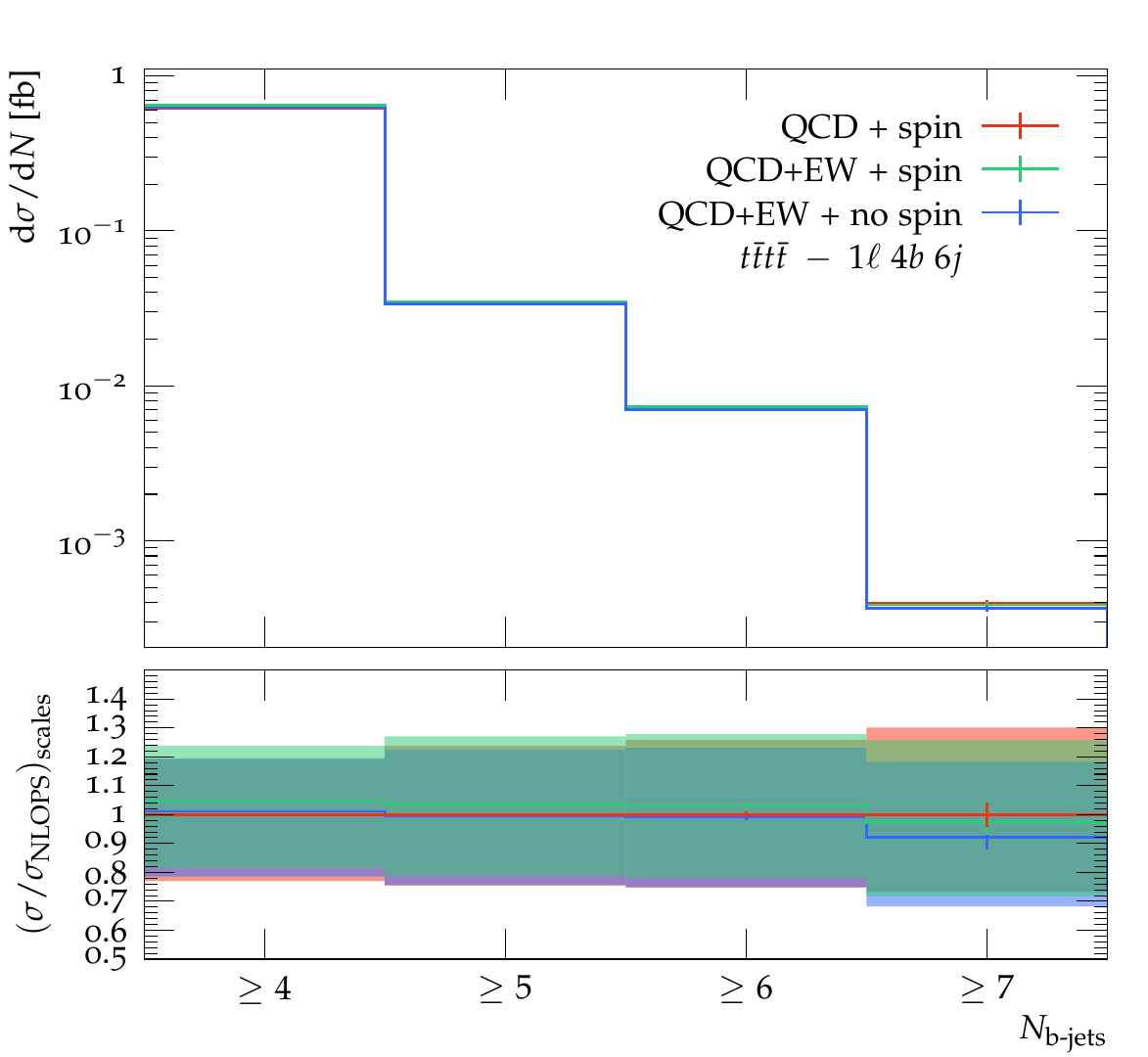}
 \caption{Inclusive cross sections in the single lepton fiducial region as a 
 function of the number of light jets (l.h.s.) and the number of $b$ jets (r.h.s.) 
 for the $pp\to t\tb t\tb$ process. The uncertainty bands correspond to 
 independent variations of the renormalization and factorization scales (bottom 
 panel).}
 \label{fig:fid_1}
\end{figure*}
We next inspect the inclusive cross section as a function of the number of light
and $b$ jets as shown in Fig.~\ref{fig:fid_1}. For both observables we recognize
essentially the same pattern, independently of the number of jets, with respect
to spin correlations and EW production modes as discussed before. However,
theoretical uncertainties grow with increasing number of jets. For instance,
uncertainties for the cross section as a function of the number of light jets are
of the order of $\pm 22\%$ for at least $6$ jets and they increase up to $\pm
36\%$ for events with at least $14$ jets. Furthermore, nearly $15\%$ of all
$pp\to t\tb t\tb$ events are accompanied by at least $9$ light jets and only
$1\%$ of all events in this signature are associated with at least $12$ light
jets.
On the contrary, the corresponding distribution as a function of the number of
$b$ jets falls off more steeply. This is expected, as the dominant source of
additional $b$ jets, besides the $4$ $b$ jets originating from the top-quark
decays, are $g\to b\bar{b}$ splittings in the parton-shower evolution.
Additionally, initial state $b\to gb$ splittings occur as part of the real
radiation contribution of the NLO QCD corrections as well as during the parton
shower evolution.  Nonetheless, these contributions are heavily suppressed by
the $b$-quark parton distribution function. For instance, only $5\%$ of all
events have one additional $b$ jet. The theoretical uncertainties as estimated
from independent scale variations start again from $\pm 22\%$ and increase up
to $\pm 30\%$ for at least $7$ $b$ jets.
Note that for more realistic estimates of perturbative uncertainties of both
cross sections as function of light and $b$ jets in bins beyond the first one
shower scale variations should be considered.

%
\begin{figure*}[ht!]
 \centering
 \includegraphics[width=0.495\textwidth]{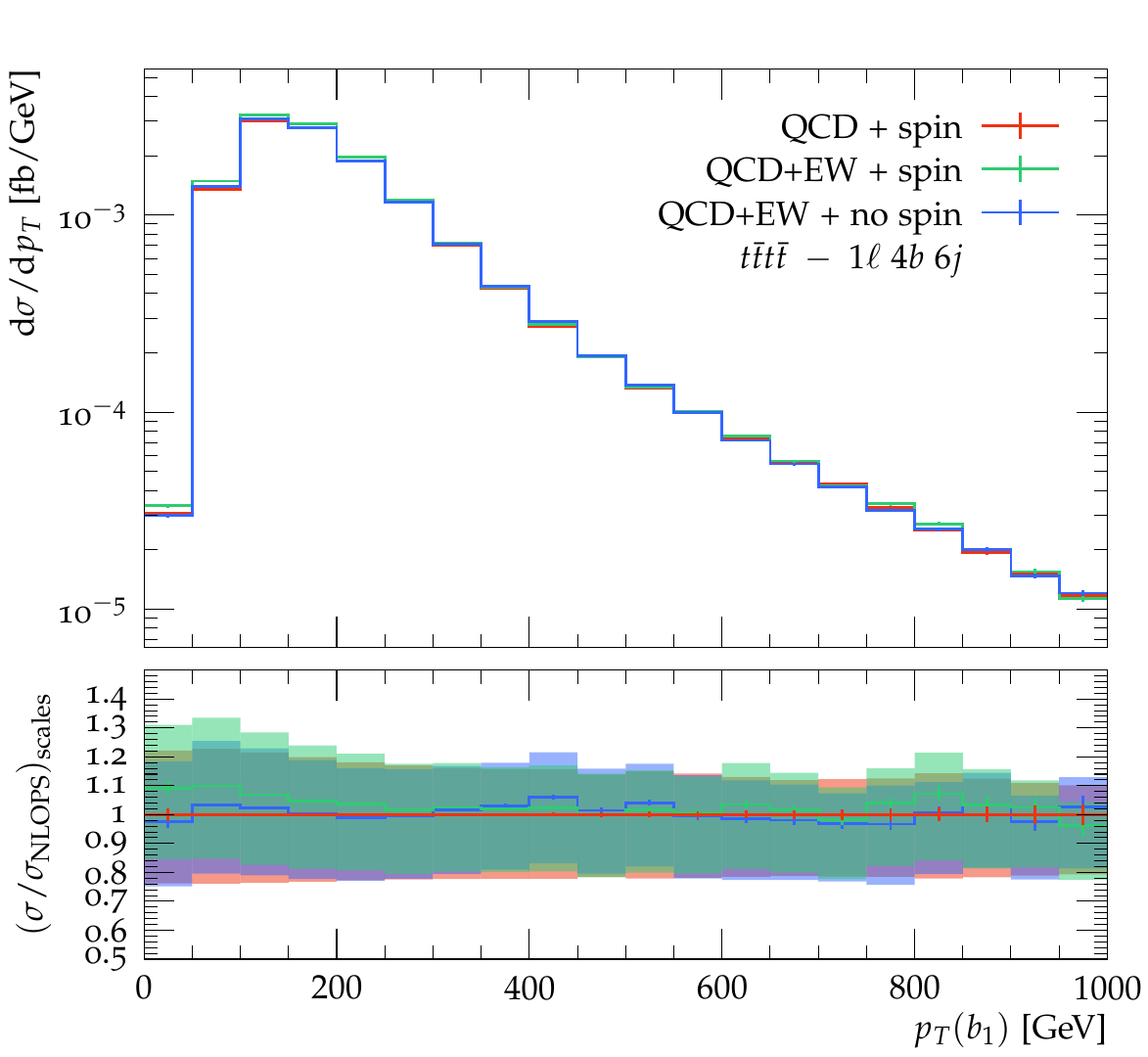}
 \includegraphics[width=0.495\textwidth]{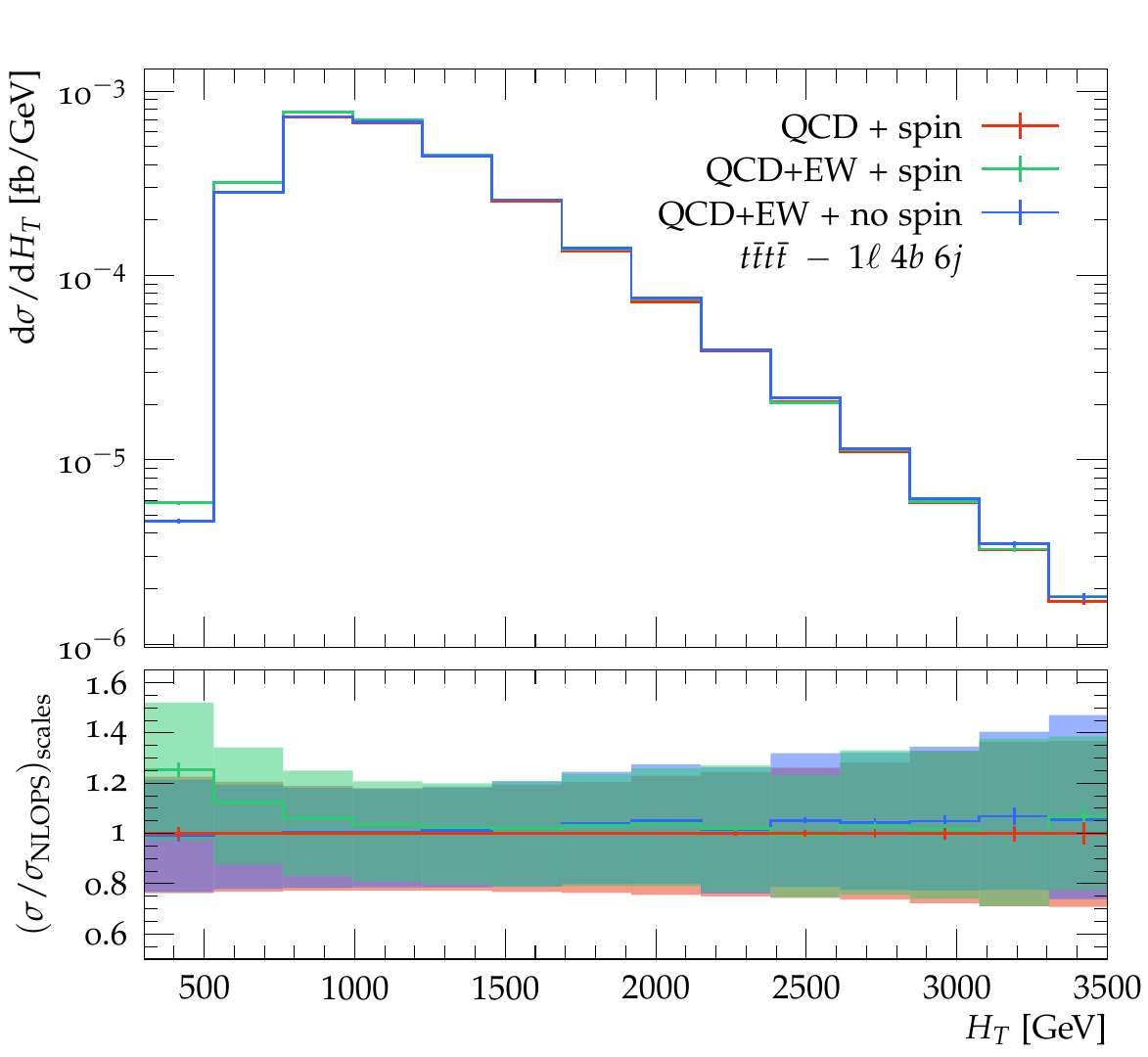}
 \caption{Differential cross section distribution in the single lepton fiducial 
 region as a function of the transverse momentum of the hardest $b$ jet (l.h.s.) 
 and of the $H_T$ observable (r.h.s.) for the $pp\to t\tb t\tb$ process.
 The uncertainty bands correspond to independent variations of the 
 renormalization and factorization scales (bottom panel).}
 \label{fig:fid_2}
\end{figure*}
Next we discuss hadronic observables such as the transverse momentum of the
hardest $b$ jet as shown in the left panel of Fig.~\ref{fig:fid_2}. First of all
we notice that the spectrum is extremely hard. Between the peak of the
distribution for transverse momenta around $150$ GeV and the tail at $1$ TeV the
cross section does not drop even $3$ orders of magnitude. Similar features have
been observed for the $p p \to t \tb b \bar{b}$ process in
Refs.~\cite{Bevilacqua:2021cit,Bevilacqua:2022twl} for final states with $4$
$b$ jets. The scale uncertainties are rather constant over the whole plotted
range and are estimated to be between $\pm 25\%$ and $\pm 20\%$, where the tail
of the distribution exhibits smaller uncertainties. Furthermore, we find a
significant contribution of up to $+10\%$ due to the inclusion of the EW
production channels for transverse momenta below $200$ GeV.

We now turn to the $H_T$ observable as depicted on the right of
Fig.~\ref{fig:fid_2}. At the fiducial level we define $H_T$ via
\begin{equation}
 H_T = p_T(\ell) + p_T^\textrm{miss} + \sum_{i=10}^{N_\textrm{jets}} p_T(j_i)\;,
\end{equation}
where we do not distinguish between light and $b$ jets. For this observable we
find an even larger impact due to the leading order EW production channels of up
to $+25\%$ below $1$ TeV. Above, they only contribute a rather constant $+3\%$
amount for the rest of the plotted range. Spin correlation effects have also only
a mild impact on the tail of the distribution.  They soften the spectrum by
roughly $5\%$. Scale uncertainties are estimated to be $\pm 22\%$ at the
beginning of the spectrum which then increases up to $\pm 35\%$ at the end of the
plotted range.

%
\begin{figure*}[ht!]
 \centering
 \includegraphics[width=0.495\textwidth]{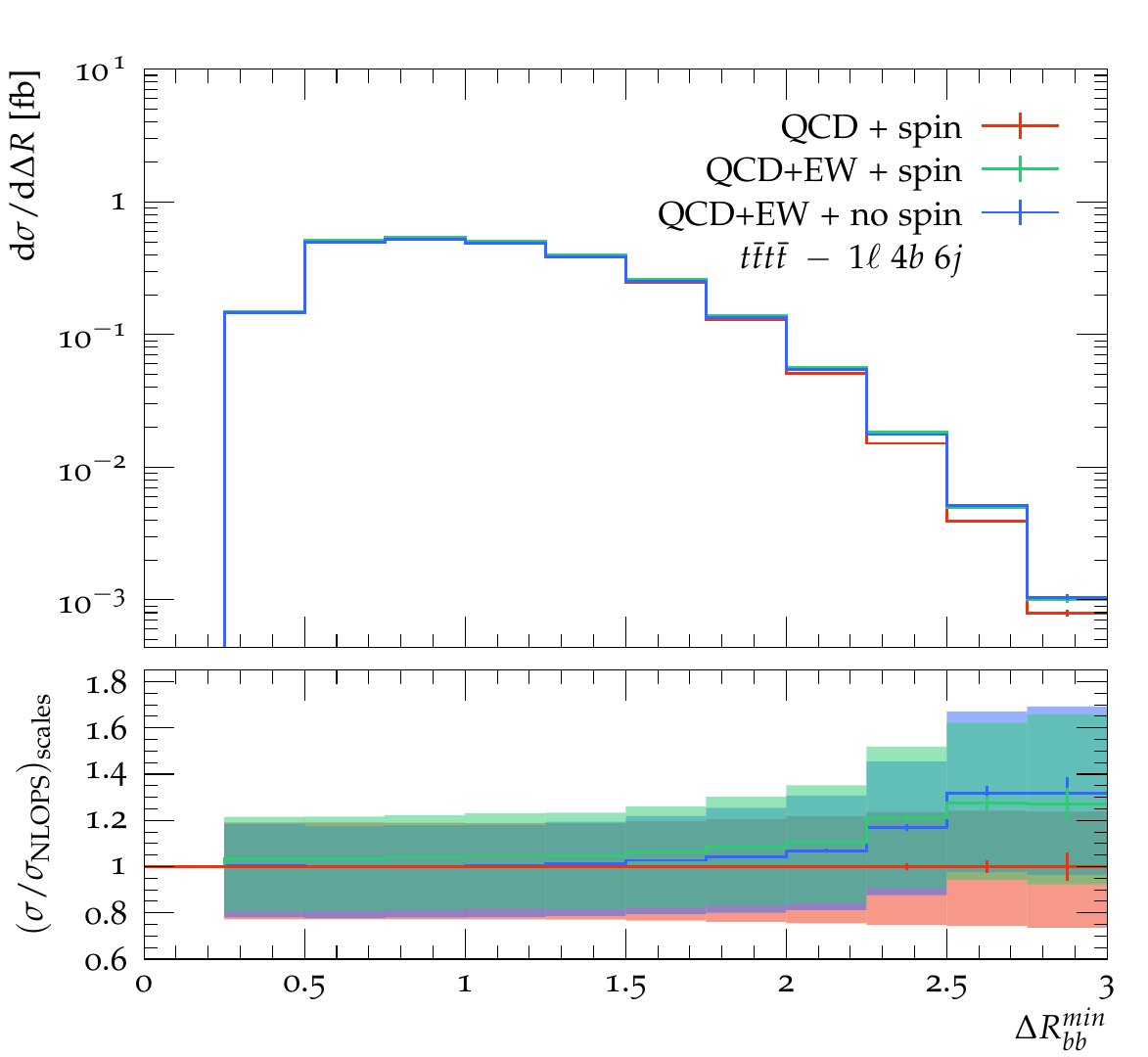}
 \includegraphics[width=0.495\textwidth]{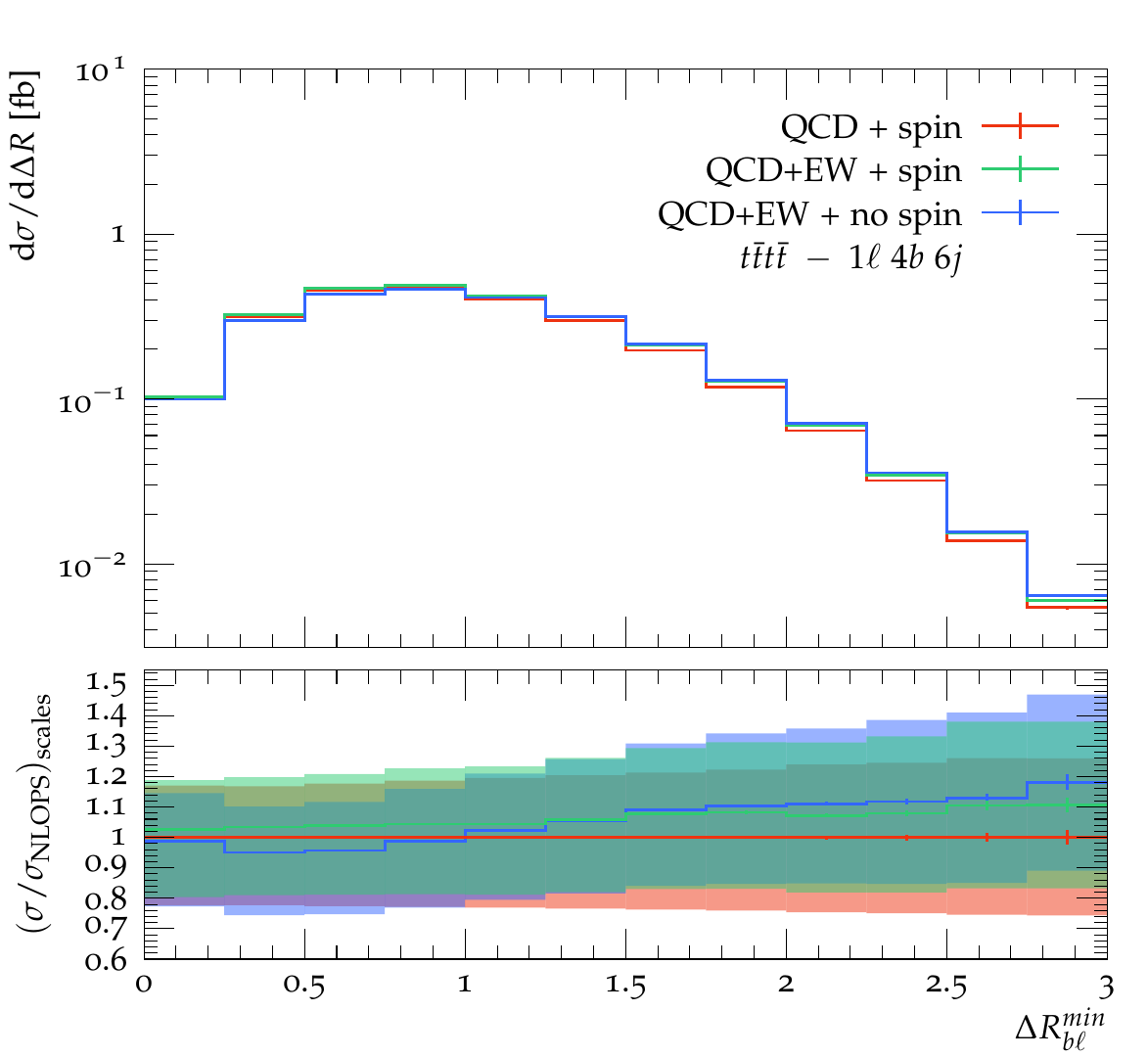}
 \caption{Differential cross section distribution in the single lepton fiducial 
 region as a function of the minimal $\Delta R$ between all $b$ jet pairs (l.h.s.) 
 and of the minimal $\Delta R$ between all $b$ jets and the lepton (r.h.s.) for 
 the $pp\to t\tb t\tb$ process. The uncertainty bands correspond to independent 
 variations of the renormalization and factorization scales (bottom panel).}
 \label{fig:fid_3}
\end{figure*}
Now we discuss two angular observables that are used for the discrimination of
the signal from the background in the experimental analysis of
Ref.~\cite{ATLAS:2021kqb}. To this end, we show in Fig.~\ref{fig:fid_3} the
minimal distance $\Delta R$ among all pairs of $b$ jets, $\Delta
R_{bb}^\textrm{min}$, as well as among all $b$ jets and the lepton, $\Delta
R_{b\ell}^\textrm{min}$.
For both angular distributions we find strong enhancements in the tails of the
distributions due to the EW production of the $pp\to t\tb t\tb$ process. To be
specific, we find corrections up to $+27\%$ for $\Delta R_{bb}^\textrm{min}$ and
up to $+11\%$ for $\Delta R_{b\ell}^\textrm{min}$. The scale uncertainties are
below $\pm 26\%$ in the whole plotted range and spin correlations only have a
very mild impact that lead to minor shape differences.

%
\begin{figure*}[ht!]
 \centering
 \includegraphics[width=0.495\textwidth]{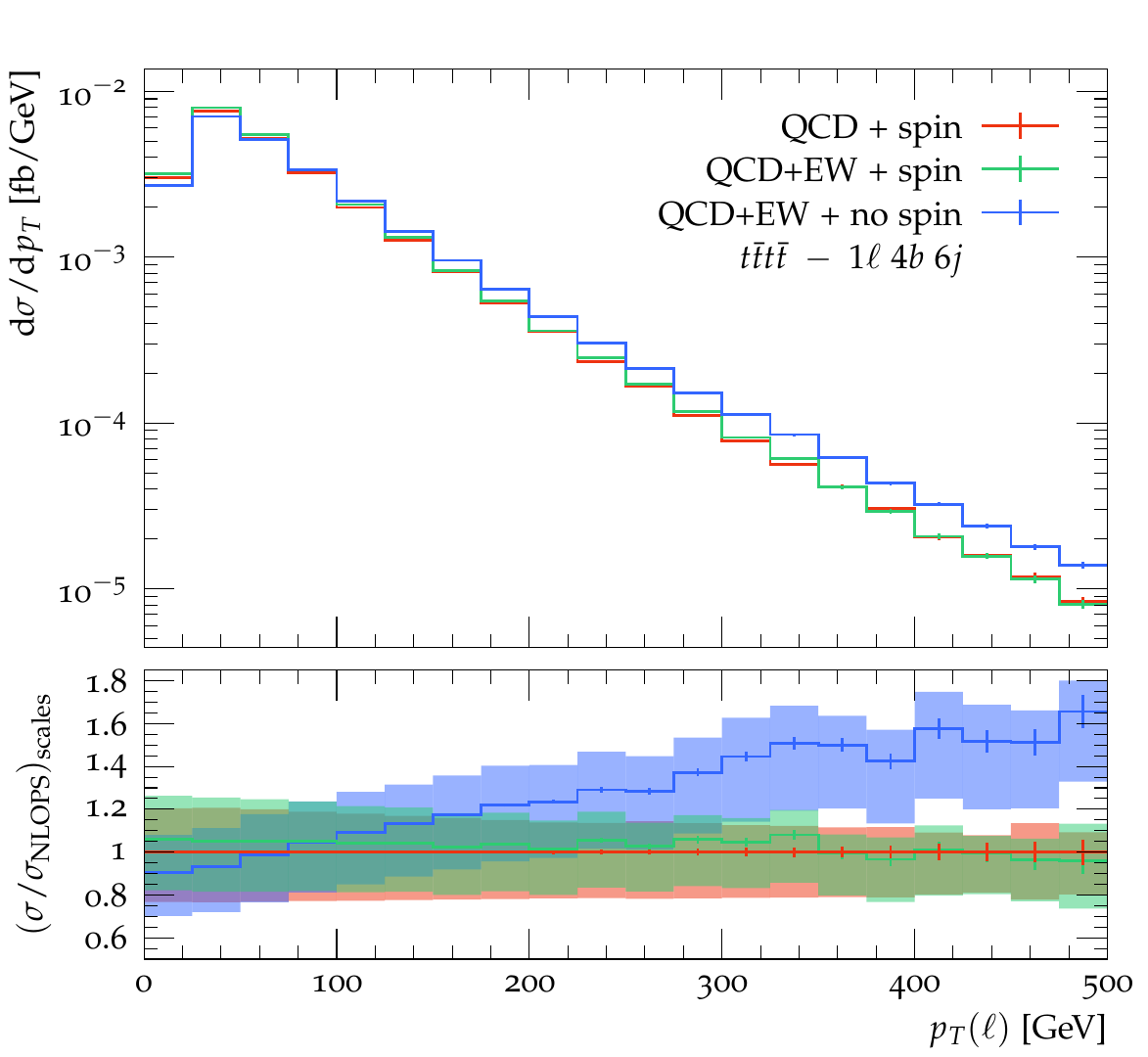}
 \includegraphics[width=0.495\textwidth]{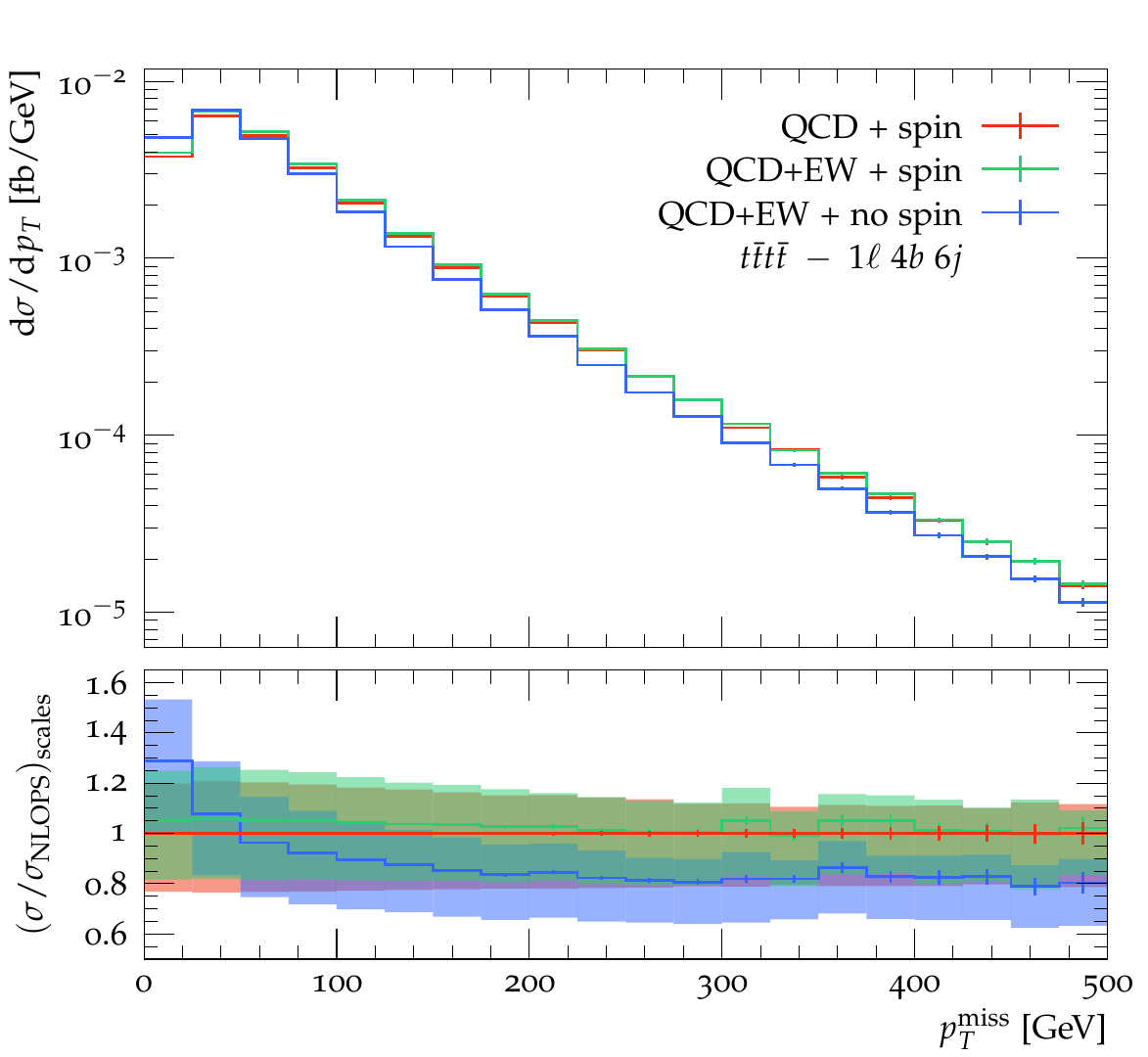}
 \caption{Differential cross section distribution in the single lepton fiducial 
 region as a function of the transverse momentum of the lepton (l.h.s.) 
 and of the missing transverse momentum (r.h.s.) for the $pp\to t\tb t\tb$ process.
 The uncertainty bands correspond to independent variations of the 
 renormalization and factorization scales (bottom panel).}
 \label{fig:fid_4}
\end{figure*}
At last we turn to leptonic observables. We show the transverse momentum of the
charged lepton on the left and the missing transverse momentum in the right panel
of Fig.~\ref{fig:fid_4}. For both observable we notice that spin correlations
have to be taken into as they have a tremendous impact on the shape of the
differential distribution. The transverse momentum distribution of the charged
lepton is much harder if correlations are omitted and overshoots the tail of the
distribution by nearly $60\%$. These deviations are not covered by the estimated
theoretical uncertainties which are of the order of $20\%-25\%$ over the whole
plotted range.

Also for the missing transverse momentum we find large shape differences.
However, in this case the distribution becomes softer as compared to the case
when spin-correlated top-quark decays are taken into account. While there are
significantly more events with low values of $p_T^\textrm{miss}$ if uncorrelated
decays are considered, for $p_T^\textrm{miss} \gtrsim 250$ GeV the spectrum is
nearly constant $-20\%$ smaller than predictions that take spin correlations into
account. Furthermore, the scale uncertainties are independent of the treatment of
decays and EW production modes of the order of $\pm 20\%$.

\section{Conclusions}
\label{sec:conclusions}
In this article we presented an implementation of the production of four top
quarks at hadron colliders in the \powhegbox{} framework. Besides taking into
account the leading NLO QCD corrections at $\mathcal{O}(\alpha_s^5)$ we also
include formally subleading EW production channels at LO accuracy. Furthermore,
our implementation allows to decay top quarks at LO accuracy retaining
spin-correlation effects. This feature made it possible to study the effects of
spin-correlations in top-quark decays in this process for the first time. 

We first investigated the impact of leading NLO QCD corrections and subleading EW
channels on the total cross sections. We find that QCD corrections contribute at
up to slightly over $+50\%$ and EW modes at the $+5\%$ to $+10\%$ level. The
inclusion of leading NLO corrections leads to a reduction of scale uncertainties
from up to $70\%$ down to at most $22\%$. Thus the inclusion of NLO QCD
corrections as well as the subleading EW channels is essential for reliable
predictions of four top production.

We investigated modeling differences for the inclusive $pp\to t\tb t\tb$
production process with stable top-quarks at the $13$ TeV LHC by comparing \mg5{}
and \powhegbox{}. We find very good overall agreement between the two frameworks
for observables at NLO accuracy, with only minor differences due to the shower
evolution in the threshold region. We do also find notable deviations in the
hardest light jet spectra, predicted at LO accuracy, which coincides with
findings in other production modes of associated top pair production. We also
investigated the impact of the EW production channels at the differential level
and estimated matching as well as scale uncertainties. We observe that the impact
of these subleading channels is generally below $10\%$ but can exceed that near
the production threshold. Nonetheless, their inclusion represents a systematic
improvement over pure NLO QCD predictions.

Furthermore, we also investigated a single lepton plus jets signature as it is
currently employed for $pp\to t\tb t\tb$ cross section measurements and addressed
for the first time the size of the subleading EW contributions at the fiducial
level. In this case they can reach up to nearly $40\%$ in distributions used for
signal/background discriminants in experimental analyses. We also studied the
impact of spin-correlated top-quark decays and found that they are essential to
obtain a reliable description of leptonic observables. We want to stress that
spin-correlation effects are indispensable also for multi-lepton signatures as
their impact is a general feature for leptonic observables and signature
independent. Finally, we also estimated theoretical uncertainties due to missing
higher-order corrections.

The $pp\to t\tb t\tb$ process still awaits its full discovery at the LHC. Once,
it is observed it will be instrumental in constraining possible new physics. We
are hopeful, that this new tool will help to study more accurately the SM
dynamics of the four top quark production process.

\section*{Acknowledgements}
The authors want to thank Silvia Ferrario Ravasio and Laura Reina for comments on
the manuscript.

The work of T.J. was supported by the DFG under grant 396021762 - TRR 257 and
through Project-ID 273811115 - SFB 1225 ``ISOQUANT'', and the Research Training
Network 2149 ``Strong and weak interactions - from hadrons to dark matter''.

M.K. thanks Fernando Febres Cordero and Laura Reina for helpful discussions. The
work of M.K. is supported in part by the U.S. Department of Energy under grant
DE-SC0010102.

The computing for this project was performed on the HPC cluster at the Research
Computing Center at the Florida State University (FSU).

\bibliography{paper}
\end{document}